# The Drake Equation as a Function of Spectral Type and Time

Jacob Haqq-Misra[1,2,*] and Ravi Kumar Kopparapu[1,2,3,4]

[1] *Blue Marble Space Institute of Science, 1001 4th Ave, Suite 3201, Seattle, WA*
[2] *NASA Astrobiology Institute's Virtual Planetary Laboratory, P.O. Box 351580, Seattle, WA*
[3] *NASA Goddard Space Flight Center, 8800 Greenbelt Road, Mail Stop 699.0 Building 34, Greenbelt, MD*
[4] *Department of Astronomy, University of Maryland, College Park, MD*

*Email: jacob@bmsis.org

**Abstract**

This chapter draws upon astronomical observations and modeling to constrain the prevalence of communicative civilizations in the galaxy. We discuss the dependence of the Drake equation parameters on the spectral type of the host star and the time since the galaxy formed, which allow us to examine trajectories for the emergence of communicative civilizations over the history of the galaxy. We suggest that the maximum lifetime of communicative civilizations depends on the spectral type of the host star, which implies that F- and G-dwarf stars are the best places to search for signs of technological intelligence today.

## 1. Introduction

The Drake equation is a probabilistic expression for the number of communicative civilizations in the galaxy (Drake, 1965). This equation typically takes the form

$$N = R_* \cdot f_p \cdot n_e \cdot f_l \cdot f_i \cdot f_c \cdot L, \tag{1}$$

where $R_*$ is the rate of star formation, $f_p$ is the fraction of stars with planets, $n_e$ is the number of habitable planets per system, $f_l$ is the fraction of habitable planets that develop life, $f_i$ is the fraction of inhabited planets that develop intelligence, $f_c$ is the fraction of planets with intelligent life that develop technology capable of interstellar communication, and $L$ is the average communicative lifetime of technological civilizations (*c.f.* Ćirković and Vukotić, 2008). The original formulation of this equation was developed by Frank Drake during a 1961 conference at Green Bank, West Virginia, with the goal of identifying the key factors needed for a planet to develop life and technology that could communicate with Earth. Since this discussion at Green Bank, the Drake equation has seen wide use among scientists and educators, with some studies proposing methods for constraining the Drake equation with observable quantities (Frank and Sullivan, 2016) or statistical methods (Maccone, 2010; Glade et al., 2012) in order to guide future search for extraterrestrial intelligence (SETI) surveys. Others have extended the traditional representation of the Drake equation to include time-dependent quantities, in an effort to understand the distribution of civilizations across the history of the galaxy (Ćirković, 2004; Glade et al., 2012; Prantzos, 2013).

In this chapter we discuss the functional dependence of the Drake equation parameters on the spectral type of the host star and the time since the galaxy formed. We discuss observations, planetary formation models, and climate models that provide constraints on the values of $R_*$, $f_p$, and $n_e$, as well as looser bounds on $f_l$ and $f_i$. We also examine common assumptions for estimating the parameter $L$ and develop two approaches for calculating the maximum value, $L_{max}$, as a function of spectral type. We use these constraints to develop an expression for $N(s,t)$, the number of communicative civilizations in the galaxy as a function of spectral type $s$ and time since galaxy formation $t$.

We also discuss and critique the possibility that $N$ has distinct phases in time, such that certain events are impossible or unlikely until a particular galactic era. Ćirković and Vukotić (2008), following Annis (1999), describe this behavior as a 'phase transition,' suggesting that phenomenon such as supernovae, gamma ray bursts, or other physical phenomena could have been more prevalent during the early phase of the galaxy compared to today. The idea of a phase transition indicates that astrophysical factors may have inhibited the emergence of communicative civilizations until relatively recently in galactic history. We argue that phase transitions for $N$ are most likely to emerge from the dependence of $L_{max}$ on stellar spectral type.

## 2. Constraints from Observations

In this section we discuss observational constraints on the parameters $R_*$, $f_p$, and $n_e$ along with their dependence on spectral type and time. We consider F-, G-, K-, and M-dwarf stars in the majority of our analysis, so that $s \in \{F, G, K, M\}$, as these present the most likely candidates for supporting intelligent life. However, casting the Drake equation as a function of spectral type also allows for consideration of O-, B-, and A-dwarf stars (which will explicitly show why such stars are deemed unlikely for advanced life) and even planets around evolved stars.

*2.1 Rate of Star Formation*

Common analyses of the Drake equation cite a value of 7 to 10 star/yr as an estimate of $R_*$. Because $R_*$ is difficult to measure directly, one way of calculating this rate is to write $R_* = n_*/t_0$, where $n_*$ is the current number of stars in the galaxy and $t_0 \approx 13$ Gyr is the current age of the galaxy. However, this uniform approach assumes that the rate of star formation has remained constant over the history of the galaxy (Ćirković, 2004), proportional to the stellar population today. As an improvement to previous estimates that offer a single value of $R_*$, we calculate values of $R_*$ that depend on spectral type.

Analysis of *Spitzer* data (Robitaille and Whitney, 2010) finds that the stellar mass formation rate is 0.68 to 1.45 $M_{sun}$/yr (where $M_{sun}$ is the mass of the sun). To convert this number into a value for $R_*(s)$, we separate the spectral types into mass ranges based on the initial mass function categories of Kroupa (2001). We assume an initial mass of 1 $M_{sun}$ for F- and G-dwarfs, 0.5 $M_{sun}$ for K-dwarfs, and 0.1 $M_{sun}$ for M-dwarfs. We then divide the observed stellar mass formation rate by this initial mass function to give $R_*(s)$, with units of number of stars formed per year. F- and G-dwarfs have the lowest rate of $R_*(F) \sim R_*(G) \sim 1$ star/yr. K-dwarfs have a rate of $R_*(K) \sim 1.3$ to 3 star/yr (we use $R_*(K) = 2.2$ star/yr in our analysis), while M-dwarfs have the highest rate of $R_*(M) \sim 7$ to 14 star/yr (we use $R_*(M) = 10.5$ star/yr in our analysis). We summarize these values for $R_*$ in Table 1.

| Star | $R_*$ | $f_p$ | $n_e$ | $f_l \cdot f_i \cdot f_c$ | $L_*$ | $L_{max}^{EET}$ | $L_{max}^{PET}$ |
|---|---|---|---|---|---|---|---|
| F-dwarf | 1 star/yr | 1 | 0.2 | 1 | 4 Gyr | 0 Gyr | 2 Gyr |
| G-dwarf | 1 star/yr | 1 | 0.2 | 1 | 10 Gyr | 6 Gyr | 5 Gyr |
| K-dwarf | 2.2 star/yr | 1 | 0.2 | 1 | 30 Gyr | 26 Gyr | 15 Gyr |
| M-dwarf | 10.5 star/yr | 1 | 0.2 | 1 | 100 Gyr | 96 Gyr | 50 Gyr |

**Table 1:** Summary of Drake equation parameters discussed in this chapter.

*2.2 Fraction of Stars with Planets*

Ground- and space-based observations all suggest that terrestrial planets are commonplace around all stars. Microlensing observations indicate that typical stellar systems contain one or more bounded planets (Cassan et al., 2012). Furthermore, analysis of *Kepler* data indicates that there are ~2 planets per cool star ($T_{eff} < 4000$ K) with periods <150 days (Morton and Swift, 2014). This orbital period encompasses the habitable zone (HZ) of cool stars. Therefore, we can safely assume that $f_p = 1$ for all spectral types F, G, K, and M (*c.f.* Frank and Sullivan, 2016).

Lineweaver et al. (2004) developed a model for the evolution of the galaxy and found that rocky planets could not have formed until about $t = 4$ Gyr, due to the lower metallicity of stars at the time. If planet formation were lower during the early history of the galaxy, then this would suggest a phase transition for the emergence of planets in the galaxy. However, these results of Lineweaver et al. (2004) are inconsistent with observations of planets orbiting stars nearly as old as the galaxy itself. The Kepler-444 system is one of the oldest known to host exoplanets, with five planets smaller than Earth orbiting this 11.2 Gyr old K-dwarf. At least for the long-lived K- and M-dwarf stars, $f_p = 1$ for $t > 1$ Gyr, which is close to the age of the galaxy. Even if F- and G-dwarf stars develop planets slightly later than this due to metallicity limitations, we remain skeptical that differences in metallicity over the history of the galaxy are responsible for a phase transition in the Drake equation. We therefore maintain the expression $f_p(s, t) = 1$.

*2.3 Number of Habitable Planets Per System*

A common approach to estimating $n_e$ is to focus on the related quantity "eta-Earth", $\eta_{earth}$, which is defined as the fraction of stars with at least one terrestrial mass/size planet within the HZ. The value of $\eta_{earth}$ can be estimated from *Kepler* data (Dressing and Charbonneau, 2013, 2015; Kopparapu, 2013; Gaidos, 2013; Petigura et al., 2013; Morton and Swift, 2014; Foreman-Mackey et al., 2014), with current estimates varying between 22% (Petigura et al., 2013) to 2% (Foreman-Mackey et al., 2014) for G- and K-dwarfs, and 20% for M-dwarf stars (Dressing and Charbonneau, 2015). In reality, $n_e$ and $\eta_{earth}$ are describing the same quantity: the average number of habitable planets (which we will assume means terrestrial planets located within the HZ) per system, is the same as the fraction of stars with at least one terrestrial planet in the HZ. This means that $n_e$ is in the range of 0.02 to 0.22 for G- and K-dwarfs, and 0.20 for M-dwarfs. Following Frank and Sullivan (2016), we will assume a constant value $n_e = 0.2$ for all spectral types.

Lineweaver et al. (2004) also argue that habitability in the early era of the galaxy would have been complicated by the intense radiation from supernova activity near the galactic core, effectively sterilizing any planets that could have formed until about 4 Gyr. However, other studies (Morrison & Gowanlock, 2015) suggest that the galactic center may be ideal locations to search for intelligent life, even with higher supernova rates. Furthermore, intense supernova activity near the galactic core does not preclude habitability at farther distances from the galactic center. Other simulations of galactic habitability suggest that the edge of a galaxy's stellar disk provides an optimal location for habitable planets during the early phase of the galaxy, while this region migrates inward at later times (Forgan et al., 2017). We therefore conclude that sterilization by supernovae at the early era of the galaxy is unlikely to result in a phase transition for $N$, so we maintain the expression $n_e(s,t) = 0.2$.

We note that gamma ray bursts have also been suggested as antithetical to habitability, particularly the suggestion that the early phase of the universe may have shown a higher rate of gamma ray bursts than observed today (Piran and Jimenez, 2014). However, other studies have argued that the radiation environment of the early universe would not have been any more intense than today, particularly for galaxies like the Milky Way, suggesting that Earth-like habitability cannot be precluded even for the early phase of the galaxy (Li and Zhang, 2015; Gowanlock, 2016).

## 3. Constraints from Theory

In this section we discuss theoretical constraints on the parameters $f_l$, $f_i$, and $f_c$ along with their dependence on spectral type and time. Other analyses (*e.g.*, Frank and Sullivan, 2016) assume that $f_l$ and $f_i$ are wholly unconstrained; however, we present insights from theoretical models of planetary habitability that can help to place bounds on these two terms. We also speculate on theoretical limits for $f_c$, although $f_c$ is much more formidable to constrain.

*3.1 Fraction of Habitable Planets that Develop Life*

The value of $f_l$ is difficult to estimate observationally, although remote characterization of exoplanet environments with spectral signatures through missions such as the James Webb Space Telescope (*JWST*) could provide one way to eventually constrain $f_l$. Spectral evidence of life on an exoplanet could be determined by observations of chemical disequilibrium, such as the simultaneous detection of $O_2$ and $CH_4$, which is maintained by biology on Earth (Des Marais et al., 2002). Some calculations with climate and photochemistry models suggest abiotic sources for $O_2$ and $O_3$ (Segura et al., 2007; Domagal-Goldman et al., 2014; Wordsworth and Pierrehumbert, 2014), although such a 'false positive' spectral signature might be discernable from inhabited planets by the detection of species such as CO and $O_4$ (Schwieterman et al., 2016). Observations by

*JWST* and future direct imaging missions will begin to provide an estimate for $f_l$ by examining the statistics of terrestrial exoplanets that show possible spectral biosignatures.

Origin of life research, both laboratory experiments and computational modeling, provides another avenue for further constraining $f_l$. This may not directly tell us about processes occurring on other planets, but origin of life research seeks to elucidate the state of possibilities that could give rise to biological phenomenon. Perhaps life can form in many possible ways other than how it occurred on Earth, and origin of life research in fields such as experimental evolution and synthetic biology (*e.g.*, Ray, 1993; Benner and Sismour, 2005; Luisi, 2016; Kacar, 2016) are potential ways of examining different pathways toward making a habitable world actually inhabited. One approach by Scharf and Cronin (2016) suggests an analytic framework, inspired by the form of the Drake equation for quantifying the probability of biogenesis by considering the availability of chemical and environmental building blocks for a given set of parameters. This particular method does not presently provide a reliable estimate of $f_l$ for planets in general, although it represents an important step toward reaching this goal.

Computational models of planetary habitability can also provide insight into $f_l$. Several authors have suggested that planets orbiting late M-dwarf stars may have experienced runaway greenhouses during the star's extended pre-main sequence phase, rendering such planets dry and uninhabitable (Ramirez and Kaltenegger, 2014; Luger and Barnes, 2015; Tian and Ida, 2015). This implies that some planets in the HZ of M-dwarf stars may be unable to develop life unless they begin with large water inventories or sufficient water is later delivered by impacts. Other calculations suggest that extreme X-ray and extreme ultraviolet radiation from active M-dwarf stars could induce hydrogen escape and further contribute to water loss on orbiting exoplanets (Airapetian et al., 2017). We cannot rule out the possibility that some of these planets will regenerate a habitable atmosphere during the star's main sequence lifetime, but this mechanism serves as an upper limit on the fraction of planets in the HZ of M-dwarf stars that could actually develop life. Although we cannot assign a number to $f_l$ based on these results, we can at least hypothesize that $f_l(M) < f_l(\{F,G,K\})$. Even if M-dwarfs can host habitable planets, they may be less likely to develop life than other spectral types due to their limited propensity to retain water.

*3.2 Fraction of Life-Bearing Planets that Develop Intelligence*

Assigning a particular value to $f_i$ can be somewhat controversial, and this value is generally considered to be unconstrained. However, insights from climate models can also assist with linking physical factors to this term.

Recent climate calculations suggest that some planets in the habitable zone may be prone to 'limit cycles' with punctuated periods of warmth lasting about 10 Myr, followed by extended periods of global glaciation lasting about 100 Myr (Kadoya and Tajika, 2014, 2015; Menou, 2015; Batalha et al., 2016; Haqq-Misra et al., 2016; Abbot, 2016). Shorter duration changes in Earth's ice coverage on timescales ranging from thousand to hundreds of thousands of years may have accelerated the pace of evolution by opening up ecological niche space for new species to occupy. Such ice age cycles are driven by long-term variations in Earth's orbit (known as 'Milankovitch cycles') or other Earth system processes (Haqq-Misra 2014). By contrast, limit cycling occurs due to changes in the rate of weathering in the carbonate-silicate cycle, on the scale of tens to hundreds of millions of years. With such prolonged 100 Myr periods of glaciation, the evolution of complex (animal-like) life could be difficult, which could therefore preclude the development of any forms of intelligence. The presence of limit cycles is a function of spectral type, due to the wavelength dependence of ice-albedo feedback, and also depends on the volcanic outgassing rate of the particular planet.

F-dwarf stars are more prone to limit cycling, but both F- and G stars show an expansion of the limit cycle region of the HZ when volcanic outgassing rates are lower than on Earth today. For example, a volcanic outgassing rate a tenth that of Earth today would reduce the HZ of F-dwarfs by about 75% and G-dwarfs by about 50% (Haqq-Misra et al., 2016). By contrast, K-type and M-type stars do not experience limit cycles at all due to the reduced effects of ice-albedo feedback. We cannot predict the presence of limit cycles based on spectral type alone, but we can write the relationship $f_i(F) < f_i(G) < f_i(\{K,M\})$. Additional modeling studies of volcanic outgassing and seafloor weathering rates expected for terrestrial exoplanets can place further constraints upon $f_i$, although observationally confirming such predictions will remain a daunting task.

Limit cycles are only one possible mechanism that could restrict the development of complex, and therefore intelligent, life on a planet located in the HZ. However, limit cycles provide a bound on $f_i$ that depends upon the net outgassing rate of the planet. This formulation links the previously unconstrained value of $f_i$ to properties of the planet itself. Other theoretical approaches may also provide insight on how the physical environment of a planet can constrain our expectations of $f_i$.

*3.3 Fraction of Intelligence-Bearing Planets that Become Communicative*

Few theoretical studies have attempted to estimate values for $f_c$, as this parameter remains extremely difficult to constrain. Anthropological research provides a potential way for approaching a value for $f_c$. The human species is incredibly diverse, and many existing indigenous people groups live in small rural communities where they depend on only a fraction of technology that the modern world requires. Although developed nations today are able to engage in spacefaring activities, it remains unclear as to how common such desires would be among intelligent beings. Until the search for extraterrestrial life succeeds, studying our fellow human beings may be the best way toward understanding the factors that drive us toward communicative technology (*e.g.*, Finney and Jones, 1986; DeVito, 2011; Denning, 2011). It remains possible that $f_c$ is dependent upon spectral type, but any such attribution remains pure speculation this point. We therefore leave $f_c$ as an unconstrained parameter.

## 4. Rethinking the Longevity Parameter

The average lifetime of a communicative civilization is perhaps the most controversial value of the Drake equation. Drake's own estimate of his equation is $N = L = 10,000$ civilizations, which suggests that the other factors multiply approximately to one (Drake, 2011). Historically, Carl

Sagan and others (Sagan, 1973) have argued that $L$ is the determining factor in the prevalence of communicative civilizations; however, much of this early speculation of the value of $L$ occurred in a geopolitical environment locked in a Cold War and facing an imminent nuclear catastrophe. Pessimism over humanity's ability to survive its own technology may have subsided in recent years, although humanity still holds the capacity to destroy itself several times over with nuclear weapons. However, we also note that $L$ represents the length of time that a civilization remains communicative, not necessarily its timescale for existence. A long-lived civilization could potentially become undetectable through the development of new technology, or perhaps even the abandonment of spacefaring technology altogether.

David Grinspoon (2004) provides an alternative and more optimistic interpretation of $L$, suggesting that long-lived civilizations will necessarily overcome any developmental challenges and become 'immortals' in the sense that they are unconstrained by existential threats. Under this interpretation, Grinspoon (2004) suggests that $L = f_{IC} \cdot T$, where $f_{IC}$ is the fraction of intelligent civilizations that become immortal and $T$ is the length of time that this process has been occurring. The value of $T$ is likely to be a fraction that may approach the age of the universe, although it is unclear how to empirically or theoretically resolve the value of $f_{IC}$.

One feature that we consider a significant omission from the Drake equation is the expected main sequence lifetime of the host star. G-dwarf stars have a typical main sequence lifetime of 10 Gyr, while F-dwarf stars evolve faster with a typical main sequence lifetime of about 4 billion years. K-dwarf stars are longer lived, with a main sequence lifetime of about 20 Gyr, while M-dwarf stars can live up to 50 to 100 Gyr or longer. The habitable lifetime of a planet depends on the evolutionary trajectory of its host star, which can be calculated with computational models. We therefore suggest a maximum value, $L_{max}$, that depends up on the evolutionary history of the star itself, as the expected lifetime of a communicative civilization cannot be any longer than the habitable lifetime of its planet.

Once a star reaches the end of its main sequence lifetime and transitions into its giant phase, a spacefaring civilization could potentially survive by migrating outward to farther regions of the stellar system (Finney and Jones, 1986). We set aside this possibility in our analysis below, limiting our consideration to main sequence F- G- and K-dwarfs that are the target of many exoplanet and SETI surveys. However, we acknowledge that any civilization able to survive into the post-main sequence phase of its host star will have a larger value of $L_{max}$ than we calculate, possibly approaching the age of the universe as with the case of Grinspoon's (2004) 'immortal' civilizations.

## 4.1 Equal Evolutionary Time

The first representation of $L_{max}$ we consider assumes that the maximum communicative lifetime is simply the difference between the host star's main sequence lifetime and the time required for the prerequisite evolutionary steps. We also note that we should expect $L_{max} = 0$ if evolutionary timescales are shorter than the host star's main sequence lifetime. Under these assumptions, we can therefore write

$$L_{max}(s,t) = \begin{cases} L_*(s) - t_{evo}, & \text{for } t > t_{evo} \\ 0 & \text{for } t < t_{evo} \end{cases}, \quad (2)$$

where $L_*$ is the main sequence lifetime of the host star and $t_{evo}$ is the average time for the evolution of intelligent and communicative life. This simplifies a formerly unconstrained parameter into one that depends upon stellar evolutionary histories. We refer to the assumptions of Eq. (2) as the *equal evolutionary time* (EET) hypothesis. EET implies that $t_{evo}$ is independent of spectral type and thus approximately constant for all environments.

EET defines $L_{max}$ in terms of the known parameter $L_*$, which we can estimate from observations and models. We are left with the unknown parameter $t_{evo}$, which is often estimated as four billion years (*e.g.*, Lineweaver et al., 2004) with Earth as our only example of communicative intelligent life. The assumption of EET with $t_{evo} = 4$ Gyr in particular implies that certain spectral types will have longer-lived communicative civilizations than others. F-dwarfs have relatively short stellar lifetimes of about 4 Gyr, which by Eq. (4) gives us $L_{max}(F) = 0$. G-dwarfs are longer-lived with a main sequence lifetime of about 10 Gyr, which gives a positive value of $L_{max}(G) = 6$ Gyr. Similarly, K-dwarfs with a 30 Gyr lifetime have $L_{max}(K) = 26$ Gyr, while M-dwarfs with a 100 Gyr lifetime have $L_{max}(M) = 96$. The logic of EET takes the timescale for the evolution of communicative life on Earth to be a static factor that occurs on average after $t_{evo}$ years, regardless of spectral type. These assumptions imply that M-dwarfs are likely to host the most long-lived civilizations because $L_{max}$ increases for later spectral types.

Various forms of EET are regularly invoked in astrobiology as a way of coping with our lack of information on evolutionary timescales in other stellar environments. However, when we include a dependence on spectral type in our version of EET, as described by Eq. (2), the overwhelming preference toward later spectral types raises questions of the suitability of a constant evolutionary timescale $t_{evo}$. We next consider an alternative formulation of $L_{max}$ that may be better suited to represent dependences on spectral type and time.

*4.2 Proportional Evolutionary Time*

Instead of assuming a constant evolutionary timescale, we can improve upon our expression for $L_{max}$ by making $t_{evo}$ dependent upon spectral type. Rather than presume that the 4 Gyr evolutionary timescale of Earth is typical, we note that humans appeared on Earth roughly halfway through the Sun's main sequence lifetime. As an alternative EET, we suggest that the relative timing of intelligent communicative life is proportional to the lifetime of the star. If we assume that communicative civilizations typically tend to arise approximately halfway through the lifetime of their host star, then $t_{evo} = L_*/2$. As a replacement to Eq. (2), we now write the expression

$$L_{max}(s,t) = \begin{cases} \frac{L_*(s)}{2}, & \text{for } t > t_{evo} \\ 0 & \text{for } t < t_{evo} \end{cases}, \quad (3)$$

which depends only on spectral type. We refer to the assumptions of Eq. (3) as the *proportional evolutionary time* (PET) hypothesis. PET implies that $t_{evo}$ depends on the main sequence lifetime of the host star and suggests that the emergence of communicative civilizations may occur at different temporal eras of the universe for each spectral type. PET suggests that F- and G-dwarfs, which are shorter lived than M-dwarfs, are more likely to host communicative civilizations today, at the present era of the universe. Eq. (3) gives $L_{max}(F) = 2$ Gyr and $L_{max}(G) = 5$ Gyr, which suggests that communicative civilizations could have already arisen around these spectral types in the history of the galaxy. However, Eq. (3) also gives $L_{max}(K) = 15$ Gyr and $L_{max}(M) = 50$ Gyr, which further suggests that communicative civilization should not yet be prominent around these later spectral types. PET predicts that M-dwarf planets are more likely to be habitable in the future than today. This could be because of physical factors, such as the high stellar activity of M-dwarfs during the early main sequence phase or the enhanced contribution of infrared radiation in the stellar spectrum—any of which could alter the temporal trajectory of evolution in such an environment. Likewise, PET suggests that the present era is probably the best time for G- dwarf stars to provide habitable conditions because the current age of the galaxy is approximately half that of the main sequence lifetime for these spectral types.

The assumptions of PET do not require that all civilizations are necessarily long-lived, and our choice of focusing on the maximum lifetime $L_{max}$ acknowledges that other factors, not necessarily linked to spectral type, could cause a civilization to achieve a value of $L$ much lower than $L_{max}$. However, such short-lived civilizations would be poor targets for SETI (Grinspoon, 2004; Haqq-Misra and Baum, 2009), so any logic that can provide upper bounds with stellar dependence on the emergence of communicative civilizations can help to constrain SETI efforts today.

## 5. Discussion

Our formulation now provides numerical constraints for $R_*$, $f_p$, $n_e$, and $L$ as well as comparative bounds for $f_l$ and $f_i$ that may be further enhanced through additional theoretical research. Only $f_c$ remains elusive, with few ties to observables or theory. We also provide expressions for $L_{max}$ that allow us to predict possible habitable trajectories over the history of the galaxy.

With the assumptions of EET, we can summarize our expression for the maximum value of the Drake equation, $N_{max}$, as:

$$N_{max}^{EET}(s,t) = \begin{cases} \frac{1}{5}R_*(s)[L_*(s) - t_{evo}]f_l f_i f_c, & \text{for } t > t_{evo} \\ 0 & \text{for } t < t_{evo} \end{cases}. \quad (4)$$

Likewise, the assumptions of PET allow us to write the maximum value of the Drake equation as:

$$N_{max}^{PET}(s,t) = \begin{cases} \frac{1}{10}R_*(s)L_*(s)f_l f_i f_c, & \text{for } t > t_{evo} \\ 0 & \text{for } t < t_{evo} \end{cases}. \quad (5)$$

It also remains possible that one of the terms $f_l$, $f_i$ or $f_c$ includes another abrupt phase transition at a later time in the history of the galaxy. Nevertheless, our expressions in Eq. (4) and (5) do represent phase transitions in the emergence of communicative civilization due to the functional dependence of $L_{max}$ on spectral type.

We proceed by assuming that the product $f_l$, $f_i$, and $f_c$ is unity, which will allow us to sketch historical trajectories for $N_{max}$ under the EET and PET hypotheses. Although we argue in Section 3.1 that $f_l$ should be lower for M-dwarfs than other spectral types due to water loss and other factors that could preclude life altogether, we also acknowledge in Section 3.2 that $f_i$ could be higher for M-dwarfs due to the reduced propensity toward limit cycling. Lacking any further observational constrains, we consider these two effects to balance each other and proceed with the assumption that $f_l \cdot f_i \cdot f_c = 1$. Our analysis is thus an optimistic case, where we make best-case-scenario assumptions regarding the emergence and longevity of communicative civilizations. Additional factors not considered in our analysis would only reduce our estimates of $N$ by lowering the value of $f_l$, $f_i$, $f_c$, or $L$. The values we use for the Drake equation parameters in each of the hypotheses is summarized in Table 1.

The temporal evolution described by Eqs. (4) and (5) are plotted in Fig. 1, which shows EET assumptions in the left panel and PET assumptions in the right panel. EET shows a single phase transition representing the origin of communicative civilizations at all spectral types (except for short-lived F-dwarfs). By contrast, PET shows phase transitions with timing that depends upon spectral type. EET suggests that all spectral types are equally likely to develop communicative civilizations, with M-dwarfs favored due to their larger number. However, PET suggests that the later era of the galaxy is better suited for the emergence of communicative civilizations around late type stars.

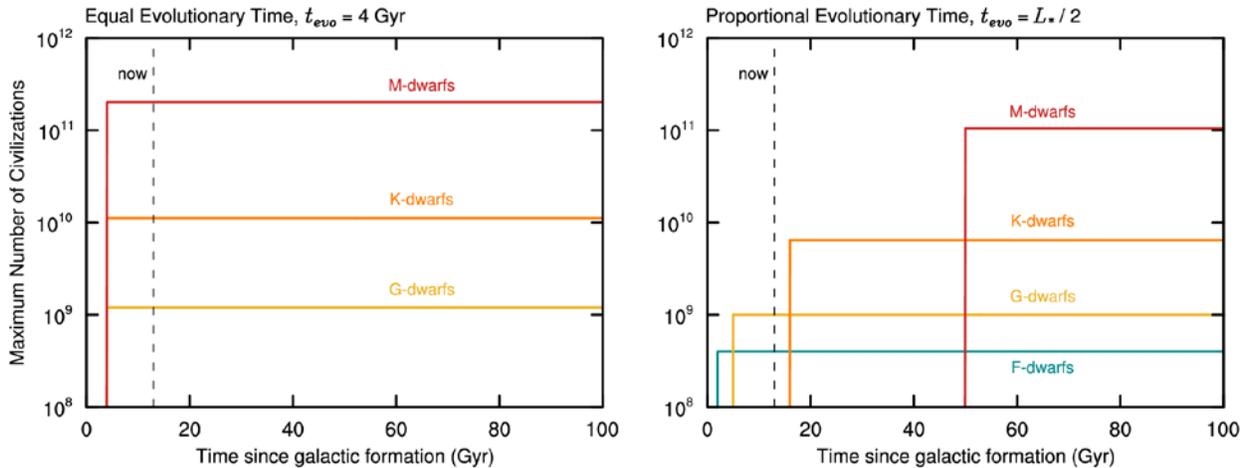

**Figure 1:** Maximum number of communicative civilizations, $N_{max}$, as a function of time since galactic formation. For the EET assumptions (left panel), communicative civilizations should be most numerous around M-dwarfs today, whereas the PET assumptions (right panel) show that communicative civilizations should be most numerous around G-dwarfs today.

The present era of the universe, under EET, should host a higher proportion of communicative civilizations around M-dwarfs compared to K- or G-dwarfs, whereas PET predicts that the present era is near the peak for communicative civilizations around G-dwarfs. Although some analyses of the galactic habitable zone make assumptions similar to EET, such that the emergence of life requires about 4 Gyr to arise on any star type (*e.g.*, Lineweaver et al., 2004), other analyses predict that the emergence of communicative civilizations depends upon spectral type (Loeb et al., 2016). We argue that the agreement between the cosmological model of Loeb et al. (2016) and our simplified Drake equation analysis with the PET hypothesis suggests that PET should be preferred over EET. More generally, even if PET is itself too limited, we recommend that other alternatives to EET should continue to be explored.

## 6. Conclusion

The galaxy has conceivably provided opportunity for other communicative civilizations to arise prior to the formation of Earth. We provide an analysis of the Drake equation that uses contemporary observations and theoretical models, arguing that none of these astrophysical parameters show obvious phase transition behavior. However, we also consider two possibilities for the timing of the emergence of communicative civilization, which does introduce phase transition behavior that depends upon spectral type of the host star.

The EET hypothesis permits the emergence of life around G-, K-, and M-dwarf stars about 4 Gyr after the formation of the galaxy. Other civilizations could conceivably have developed between this period and the formation of Earth (about 9 Gyr after the galaxy formed). This 5 Gyr period in the history of the galaxy provides a wide window for the development of biology, intelligence, and communicative civilization on other worlds. Unless other factors contribute to a more recent phase transition for the emergence of intelligence, EET suggests a history where previous civilizations in the galaxy may have risen and fallen (or perhaps even still exist today). EET suggests that G-, K-, and M-dwarfs would make good targets for SETI today, but M-dwarfs should be preferred because they have a greater value of $N_{max}$.

By contrast, the PET hypothesis permits the emergence of life around all spectral types, with a timing that occurs halfway through the main sequence lifetime of the host star. PET predicts that F-dwarfs could develop communicative civilizations as early as 2 Gyr after the galaxy formed, while G-dwarfs civilizations would have emerged at 5 Gyr. This again provides a window of about 7 Gyr between the first F-dwarf civilizations and the formation of Earth. PET predicts an even greater historical window than EET for the past emergence of communicative civilization. PET also predicts that the emergence of communicative civilization on K- and M-dwarfs remains in the future, so G- and possibly F-dwarfs are the best targets for SETI today.

Although both EET and PET make simplifying assumptions, we argue that PET is preferable for guiding SETI observations. The conspicuous absence of signals detected by SETI thus far suggests that communicative civilizations are rare enough that a much more prolonged search is required—or else, perhaps our detection methods are flawed. EET would indeed suggest that the silence of SETI means that intelligence is rare and we are alone—after all, if life can arise with equal timing regardless of spectral type (and if technological intelligence is common), then we should expect to soon see signs of communicative civilization on M-dwarf systems with all-sky surveys such as the Breakthrough Listen initiative. Instead, PET predicts that the peak of civilizations for K- and M-dwarfs remains in the future, so the lack of signals from these type of systems is to be expected. PET suggests that G-dwarf systems, perhaps some older than our solar system, are the best targets to search for signs of intelligent communicative life today.